\documentclass[12pt,preprint]{aastex}

\slugcomment{Accepted by The Astrophysical Journal}
\shorttitle{Companion Limits for Fomalhaut and $\epsilon$ Eridani}
\shortauthors{Marengo et al.}

\def\epseri{$\epsilon$~Eridani}
\def\spitzer{\textit{Spitzer}}
\def\asec{$^{\prime\prime}$}
\def\amin{$^{\prime}$}

\def\deg{$^{\circ}$}
\def\um{$\mu$m}

\begin{document}

\title{Spitzer/IRAC Limits to Planetary Companions of  \\
Fomalhaut and $\epsilon$ Eridani}

\author{Massimo Marengo}
\affil{Harvard-Smithsonian Center for Astrophysics, 60 Garden St,
 Cambridge, MA 02138 USA}
\email{mmarengo@cfa.harvard.edu}

\author{Karl Stapelfeldt, Michael W. Werner}
\affil{Jet Propulsion Laboratory, California Institute of Technology,
 4800 Oak Grove Drive, Pasadena CA 91109 USA}

\author{Joseph. L. Hora, Giovanni G. Fazio, Michael T. Schuster}
\affil{Harvard-Smithsonian Center for Astrophysics, 60 Garden St,
 Cambridge, MA 02138 USA}

\author{Joseph C. Carson}
\affil{Max-Planck-Institut f\"ur Astronomie, K\"onigsthul 17, 69117
 Heidelberg Germany} 

\author{S. Thomas Megeath}
\affil{The University of Toledo, 2801 West Bancroft Street, Toledo,
 OH 43606 USA}

\begin{abstract}

  Fomalhaut and \epseri{} are two young, nearby stars that possess
  extended debris disks whose structures suggest the presence of
  perturbing planetary objects.  With its high sensitivity and stable
  point spread function, \spitzer{}/IRAC is uniquely capable of detecting
  cool, Jupiter-like planetary companions whose peak emission is
  predicted to occur near 4.5~\um. We report on deep IRAC imaging of
  these two stars, taken at 3.6 and 4.5~\um{} using subarray mode and
  in all four channels in wider-field full array mode.  Observations
  acquired at two different telescope roll angles allowed faint
  surrounding objects to be separated from the stellar diffraction
  pattern.  No companion candidates were detected at the reported
  position of Fomalhaut b with 3$\sigma$ model-dependent mass upper
  limits of 3~M$_{J}$ (for an age of 200~Myr). Around \epseri{} we
  instead set a limit of 4 and $\la 1$~M$_J$ (1~Gyr model age) at the
  inner and outer edge of the sub-millimeter debris ring,
  respectively. These results are consistent with non-detections in
  recent near-infrared imaging searches, and set the strongest limits
  to date on the presence of planets outside \epseri{} sub-millimeter
  ring.

\end{abstract}



\section{Introduction}\label{sec-intro}

Debris disks are dust clouds in extrasolar planetary systems produced
by ongoing collisions between small bodies analogous to asteroids and
comets.  Their internal structure is of considerable scientific
interest, for it can directly betray the presence of planets.  Narrow
rings, warps, central holes, and azimuthal asymmetries can be related
to the likely mass and orbital characteristics of the perturbing
object \citep{Liou99, Wyatt99}. Nearby debris disks are thus ideal
targets for deep imaging searches for extrasolar planets.

Fomalhaut ($\alpha$ PsA; IRAS 22549-2953) is a bright A3V star 7.7~pc
distant, with an estimated age between 100 and 300~Myr
\citep{Barrado98}.  An inclined circumstellar dust ring, with an inner
and outer edge diameter of 260 and 310 AU respectively, was first
resolved by \citet{Holland98} at 850 \micron.  A modest ring
brightness asymmetry seen at 450 \micron{} \citep{Holland03} becomes
increasingly prominent toward shorter wavelengths in {\it Spitzer
  Space Telescope} images \citep{Stapelfeldt04}, consistent with
models of an eccentric ring that is warmed at periastron.  This model
was subsequently confirmed in detail by HST scattered light imaging
\citep{Kalas05}.  The sharpness of the ring inner edge is consistent
with dynamical sculpting by a giant planet orbiting nearby
\citep{Quillen06}, recently detected by \citet{Kalas08} at optical
wavelengths.

$\epsilon$ Eridani (IRAS 03305-0937) is a K2V star only 3.2~pc
distant, with an estimated age of $\sim 850$~Myr \citep{Difolco04}.  A
submillimeter dust ring, 220 AU in diameter and projected close to
face-on, was discovered by \citet{Greaves98}.  Azimuthal clumps seen
along the ring have been suggested to be dust trapped in orbital
resonances with a giant planet \citep{Ozernoy00, Quillen02}, but the
brightness and location of these clumps have not been reproducible
\citep{Schutz04, Greaves05, Backman09}.  Far-infrared imaging and
spectroscopy with {\it Spitzer} indicate the presence of two interior
dust rings, with additional planets suggested to clear the radial gaps
in the dust distribution \citep{Backman09}.  The star has been the
target of many near-infrared companion searches \citep{Macintosh03,
  Carson05, Nakajima05, Itoh06, Biller07, Janson07, Janson08,
  Heinze08, Kenworthy09}.  Several of these studies focused on
detecting the radial velocity planet that that orbits about 1\asec{}
from the star \citep{Hatzes00}, but without success.

Models for the emergent spectra of young giant planets show a
prominent emission peak near 4.5 $\mu$m, caused by a gap in CH$_4$,
H$_2$O and NH$_3$ opacity \citep{Burrows04, Fortney08}. \spitzer{}/IRAC
is well-suited to searching for such objects, which in broad-band
photometry will appear bright in channel 2 and much fainter in channel
1 \citep{Patten06}.  T dwarf companions to two nearby stars were
discovered in this way by \citet{Luhman07}.  A deep IRAC search for
planetary companions to $\epsilon$ Eridani was conducted by
\citet{Marengo06}.  No companion candidates were found to a
model-dependent mass limit of 1 Jupiter mass, in the region $r\ga$
25\asec{} (80 AU).  Inside this radius, the imaging sensitivity was
compromised by bright star artifacts.

The IRAC subarray imaging mode offers rapid readouts and exposure
times as short as 0.02~s.  To probe the 5-20\asec{} region where the
Fomalhaut and $\epsilon$ Eridani debris rings are found (and where
planetary companions perturbing the rings are expected), we conducted
a program of deep IRAC subarray imaging of these two stars.  In this
paper, we present the results of our IRAC subarray imaging search, a
new analysis of the wider-field IRAC full array dataset, limits to the
surface brightness of the debris disks at IRAC wavelengths, and limits
to the mid-infrared brightness of Fomalhaut b and of planets in
proximity of the \epseri{} rings.

\section{Observations and data reduction}\label{sec-obs}

The observations are summarized in Table~\ref{tab-obs}.  The IRAC
subarray observations were carried out as a part of \spitzer{} General
Observer (GO) program 30754 using only band 1 (3.6~\um) and band 2
(4.5~\um). Each observation requested 255 repeated stacks of $64
\times 0.02$~s frames (0.01~s integration time) on 9 dither positions,
spaced over the subarray on a small-scale Reuleaux triangle. The
choice of the frame time and dither pattern was guided by the need to
reduce saturation as much as possible and provide the best image
sampling on the under-sampled IRAC pixel grid, in order to allow the
removal of the stellar Point Spread Function (PSF). Fomalhaut was
observed with this sequence on 2006 November 24 and 2006 December 26,
with a 11.67\deg{} clockwise roll angle offset between the two epochs.
\epseri{} was observed on 2007 February 16 and 2007 September 13,
providing a 169.03\deg{} roll angle offset. The total on-source
integration time for each sequence of $9 \times 225 \times 64 =
146,880$ frames was 1468.8~s. The first Fomalhaut sequence executed on
2006 November 24, however, was cut short due to a bug in the
\spitzer{} time allocation software, skipping the last 77 repeats in
the final dither position while observing in band 2. The integration
time for the Fomalhaut first epoch 4.5~\um{} observation is thus only
1,419.5~s. The total field of view for the subarray observations is
$\sim 44$\asec.

IRAC full array observations of Fomalhaut took place on 2004 Oct 31
and 2004 Nov 28, as a part of Guaranteed Time Observer (GTO) program
90.  The roll angle offset between the two epochs was
11.03\deg{} 
clockwise.  At each epoch, the observations consisted of four 12~s
frames (10.4~s integration time) taken at each of 12 dither positions
spaced over the array on a small-scale Reuleaux triangle. The total
integration time was 499.2~s in each of the four IRAC bands.  The
total field of view imaged in each band was $\sim 5.78$\amin. 
\epseri{} was observed on 2004 Jan 9 and 2004 Feb 17 with a roll angle 
offset of 20.15\deg{} clockwise. The total integration was of 3369.6~sec 
in each of the 4 IRAC bands, obtained using individual 12~sec frame-time 
observations on a 36 points Reuleaux small scale dither pattern, repeated 
9 times in each position.  The \epseri{} full frame IRAC data has been
previously presented in \citet{Marengo06}.
We include here a new analysis of this data to take advantage of the latest 
photometric calibration of IRAC, and to ensure a consistent data reduction 
of the full frame images with the subarray observations.

The two stars, in both subarray and full frame observations, are
saturated. While the full frame data are saturated within a radius of
$\sim 6$\asec{} from the star, the much shorter frame time of the
subarray observations restricts saturation to the central 1-2 IRAC
pixels ($\sim 1.5$--3\asec{} radius). This allows to search for low
mass companions in the subarray data at closer distances from the
star than in the full frame data.

\subsection{Data Reduction}\label{ssec-data-reduction}

Basic data reduction for all the observations was performed with the
\spitzer{} Space Center (SSC) IRAC Pipeline version S15, which
produced Basic Calibrated Data (BCD) frames and data quality masks for
each individual full frame and subarray exposure. The full frame data
were reduced using the IRACproc package \citep{Schuster06}, to obtain
a single flux calibrated mosaic combining all the individual exposures
in each epoch and each band, on a pixel grid with 0.24\asec/pixel
resolution. IRACproc is based on the SSC mosaic software MOPEX and
provides enhanced outlier (cosmic rays) rejection.

We reduced the subarray data with our custom software, due to the very
large number of frames acquired for each exposure, and to mitigate
strong ``muxstripe'' artifacts in the BCDs. The muxstripe artifacts
appear as a ``jailbar'' pattern repeating with a 4 columns cadence, and
is triggered in the InSb arrays (bands 1 and 2) by the flux of bright
sources unbalancing the 4 multiplexer readout channels. The striping
pattern spans the whole array, but is stronger below the triggering
source, where an exponential decay pattern is superimposed with the
same 4 column cadence. The muxstripe is poorly characterized (does
not scale linearly with the source brightness) and its strength
depends on the Fowler sampling numbers of the exposures. We
unexpectedly found that for the brightness of our sources and the
chosen frame time the muxstripe artifacts were dominating all other
sources of noise (as much as $\sim 50$ times above the expected
read-out noise in the final mosaics).

\begin{figure}[t]
\begin{center}
\includegraphics[width=0.85\textwidth]{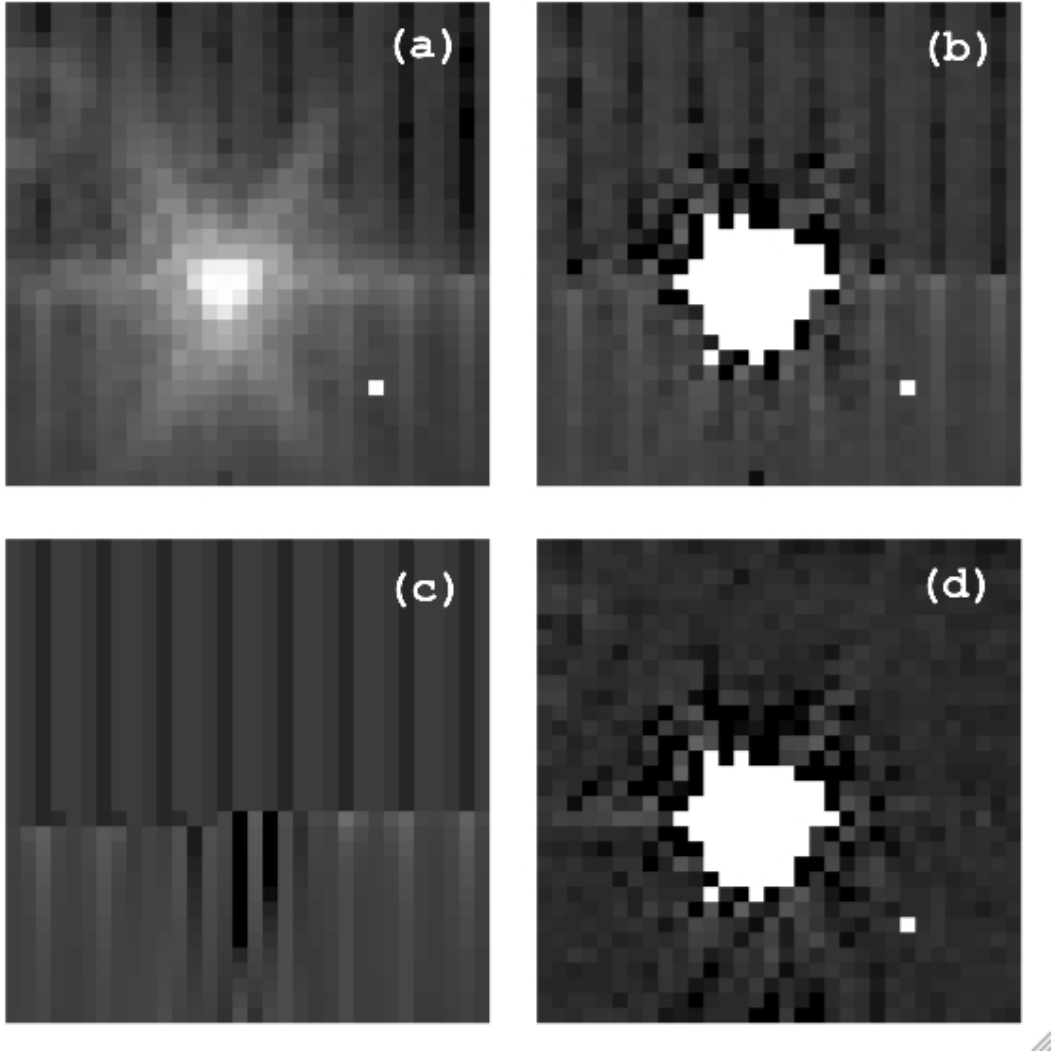}
\end{center}
\caption{Correction of the muxstripe artifacts for the first dither
 position of the 2007 February 16 \epseri{} observations: \emph{(a)}
 coadded stack of $256 \times 64$ subarray images; \emph{(b)} the
 same stack after PSF subtraction; \emph{(c)} muxstripe correction
 matrix; \emph{(d)} final PSF-subtracted stack with correction
 applied. The white area at the center of the (b) and (d) panels has
 been masked to exclude pixels where the muxstripe artifacts cannot be
 efficiently corrected.}\label{fig-muxstripe}
\end{figure}

An example of the problem is shown in Figure~\ref{fig-muxstripe} for a
stack of $255 \times 64$ co-added (with a basic temporal outlier
filter) frames (panel \emph{a}). The first step we took to
characterize the artifacts was to isolate them by subtracting the
stellar PSF, which we constructed by combining the subarray
observations of 25 bright Cepheid stars obtained as part of the
General Observer program 30666. These Cepheids are bright enough to
provide a high S/N PSF without being saturated. Observed with frame
times of 0.1 and 0.4~s, they still suffer from muxstriping, but at a
much lower level than our 0.02~s frames, and without the presence of
the exponential decay. The ``Cepheid PSF'' was aligned with each $255
\times 64$ co-added image using the diffraction spikes (typically with
a precision of better than $\sim 0.2$~\asec) and
subtracted. Figure~\ref{fig-muxstripe} (panel \emph{b}) shows the same
frame after PSF subtraction. While the PSF structures are not fully
removed (due to the different observing parameters and source colors),
the muxstriping is more clearly seen. To characterize this artifact we
first equalized the column offset pedestal between the 4 multiplexer
readout channels. We then fitted each column separately, in the lower
part of the image, with either an exponential function of the form $y
= b \cdot \exp(-x/a) + c$ or a linear slope $y = b x + c$ (in case of
not converging exponential fit). The result of this operation is a
``correction matrix'' (panel \emph{c} in Fig~\ref{fig-muxstripe}) that
is then applied to the PSF-subtracted $255 \times 64$ co-added stack
(panel \emph{d}). The correction was derived independently for each of
the 9 dither positions, as the artifacts depend on the position of the
star on the pixel grid.

The 9 individual images obtained for each dither position with the
procedure described above were then aligned (using the centroids
derived when removing the ``Cepheid PSF'') and co-added to produce one
image per band, and per epoch, for each source.

\subsection{PSF roll-subtraction}\label{ssec-psfsub}

We have combined the data for the two epochs in one single image for
the full frame and subarray observations, separately. By taking
advantage of the different roll angle, we have removed the stellar
PSF while preserving the other point sources in the field. 

\begin{figure}[t]
\begin{center}
\includegraphics[width=0.9\textwidth]{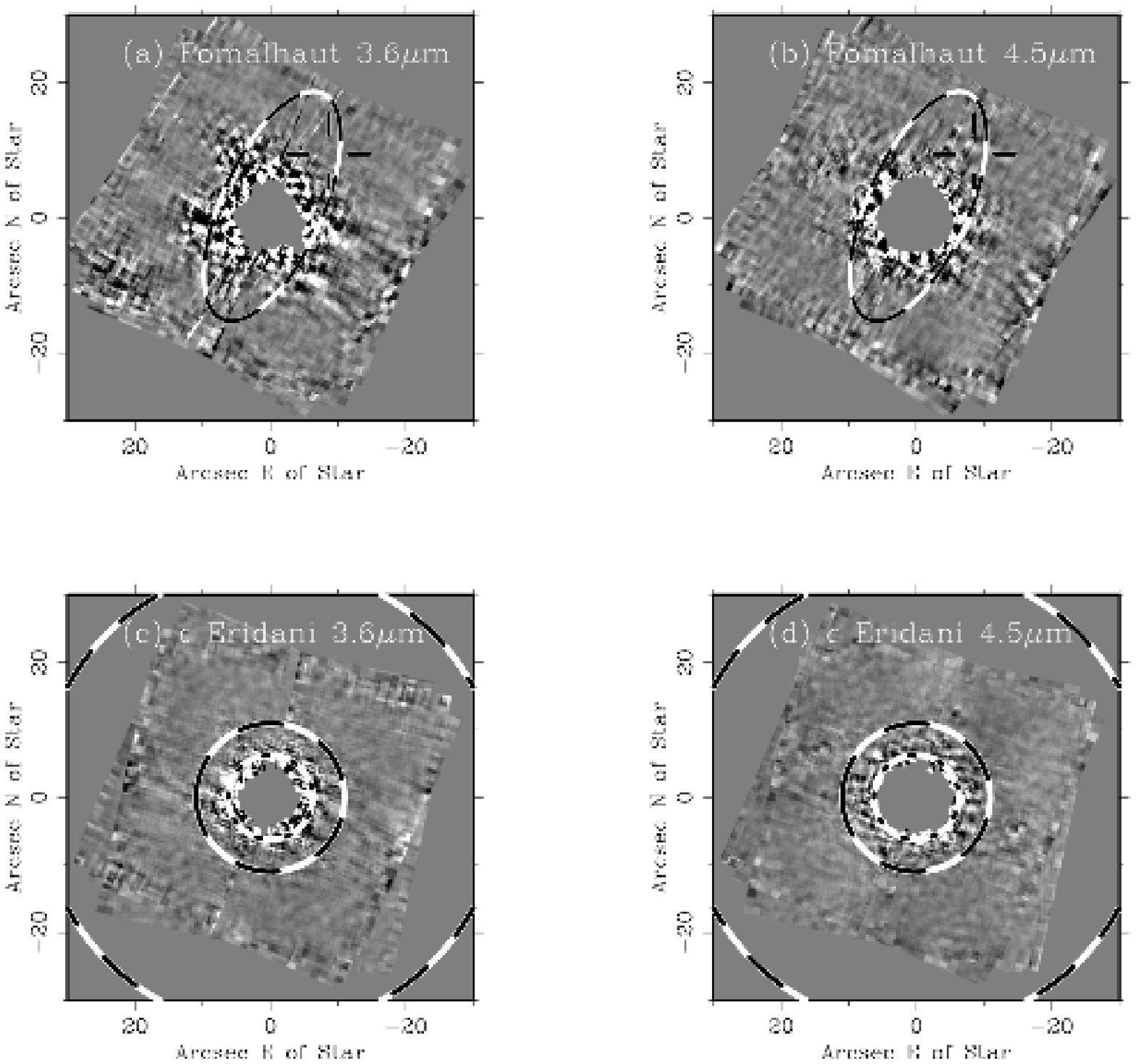}
\end{center}
\caption{Final roll-subtracted subarray images: (a) Fomalhaut
  3.6~\micron, (b) Fomalhaut 4.5~\micron, (c) \epseri{} 3.6~\micron{}
  and (d) \epseri{} 4.5~\micron. The images are scaled linearly from
  $-20$ to $+20$ MJy/sr. The ellipse and cross in panels (a) and (b)
  indicate the position of the debris ring and of Fomalhaut b. The
  dashed circles in panels (c) and (d) indicate the inner and outer
  radii of the \epseri{} sub-millimeter ring while the dotted circle
  shows the location of the 20~AU ``asteroid belt'' (from
  \citealt{Backman09}). The central region of each image is masked
  (because of saturation and PSF subtraction artifacts). The bright
  spot NE of \epseri{} is most likely a PSF subtraction artifact,
  being too narrow (less than 1/2 of the PSF FWHM) and lacking the
  two negative aliases expected for real astronomical point
  sources in the PSF roll-subtracted images.}\label{fig-subarray}
\end{figure}

In the case of the subarray data we aligned the two images obtained
for each epoch. Even though these images have been already
PSF-subtracted with the ``Cepheid PSF'', this subtraction leaves
significant residuals because of the different dither pattern
adopted. To remove these residuals, we subtracted the aligned images
one with the other. After masking in each image the row centered with
the subtracted star (where the muxstripe correction is less accurate),
we then rotated the subtracted images to have N up and E left. We then
averaged them to obtain one single image for each source, in each band
(Figure~\ref{fig-subarray}). Because of this procedure, point sources
in the field of these images would appear as one bright spot sided by
two negative sources having half height: inspection of the images in
Figure~\ref{fig-subarray} clearly shows that we do not detect any
source within the subarray field of view towards either stars.

In the case of the full frame images, we have taken advantage of the
availability of a high S/N IRAC PSF to improve the PSF removal beyond
what is achievable with the standard 2 epochs roll-subtraction
technique. We derived the full frame IRAC PSF by combining
observations of all the stars in the GTO program 90 (2 epoch images of
\epseri, Fomalhaut and Vega, 1 epoch image of $\epsilon$~Indi) with
the exclusion of $\beta$~Pic (because of the presence of its debris
disk, detected at IRAC wavelengths), plus an image of Sirius obtained
as part of the IRAC calibration program. We first reduced the
individual PSF star images using IRACproc, producing mosaics on a
pixel grid rotated as the IRAC arrays (to preserve the orientation of
the PSF features) with a pixel scale of $\sim 0.24$\asec/pixel. We
masked any pixel with flux higher than 80\% of the IRAC saturation
limit, to ensure linearity. We then aligned all the images together,
using the diffraction spikes as reference (this typically provides an
accuracy of $\sim 1/10$ of the pixel grid, or $\sim
0.02$\asec). Finally, we rescaled all the images to the same reference
(one of the images of Vega) and then co-added them together using a
median filter to reject all point sources in the field. This produced
a single PSF image in each band, calibrated to have the same surface
brightness (in MJy/sr) as Vega, cleaned of all point sources in the
field. Note that, as a byproduct of this procedure, the normalization
factor of the PSF stars are nothing else than the flux ratios of these
stars with Vega. These ratios can be express as a measure of the Vega
magnitudes of the stars, completely independent from the absolute IRAC
calibration (because they are the ratios of IRAC images of the stars
with an actual image of Vega), and are listed in
Table~\ref{tab-psf-phot}. The typical accuracy of these magnitudes is
$\sim 0.01$~mag. These PSFs are available at the SSC web
site\footnote{http://ssc.spitzer.caltech.edu/irac/psf.html}.

The procedure described above is similar to the one adopted in
\citet{Marengo06}, but results in a more accurate PSF because of the
addition of Sirius and because of a better masking procedure,
preserving linearity closer to the center (ensuring more accurate PSF
subtractions near the star, which is crucial for this program). While
this PSF has a very high S/N and is cleaned of the point sources in
the field, it still suffer from uncorrected electronic artifacts
(column pull-down, muxbleed, bandwidth effect and banding). These
artifacts are not linearly dependent on the source flux, and are not
removed by simple PSF-subtraction. To improve this situation, we have
derived two specialized version of the PSF, suitable to remove both
electronic and optical PSF features, one optimized for Fomalhaut, and
the other for \epseri.

To achieve this, we first aligned and rescaled the two images of each
source (one for each epoch, left in the IRAC array orientation, so
that the structures of the PSF will overlap), with the PSF. We then
median combined the three-image stack, obtaining a new PSF that: (1)
in areas free of electronic artifacts, it is the median of the
original PSF with the two epoch images; and (2) where electronic
artifacts occur, it samples values of the artifacts from one of the
two source images. This optimized PSF is free of background sources
(which are median filtered) and at the same time has the electronic
artifacts scaled appropriately for the source fluency.  We then
aligned and subtracted the optimized PSF from each of the two epochs,
achieving a subtraction of both the optical and electronic structures
of the PSF. This subtraction is of higher quality that the standard
two epochs roll-subtraction, because the optimized PSFs are free from
background sources.

\begin{figure}[t]
\begin{center}
\includegraphics[width=0.80\textwidth]{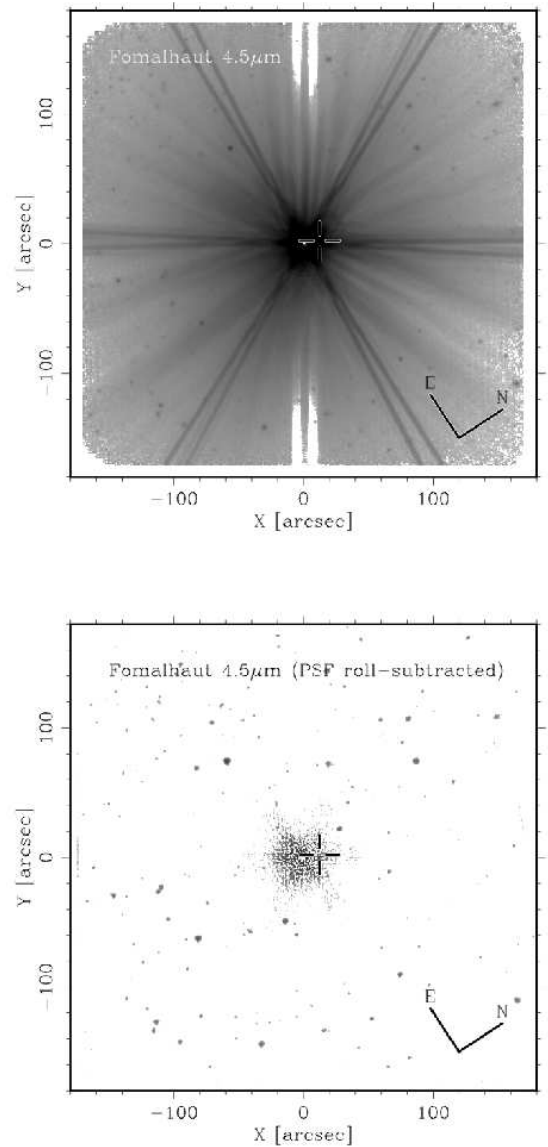}
\end{center}
\caption{Example of PSF roll-subtraction with IRAC full array
  data. The top panel shows the Fomalhaut first epoch image
  (4.5~\micron). The bottom panel shows the final two-epochs
  roll-subtracted image of Fomalhaut at the same wavelength. Both
  images are shown in logarithmic scale, from 0 to 100~MJy/sr in the
  surface brightness color scale. The cross indicates the position of
  Fomalhaut b. An enlarged figure of the central area around Fomalhaut
  and \epseri{} is shown in
  Section~\ref{ssec-photometry}.}\label{fig-psfsub}
\end{figure}

After subtracting each of the two epoch images from their optimized
PSF, we projected the result to have N up and E left, and then
averaged the rotated images. A sample PSF roll-subtracted image is
shown in Figure~\ref{fig-psfsub} (bottom panel, the top panel shows
the image before PSF subtraction) for Fomalhaut at
4.5~\micron. Figures of the final PSF roll-subtracted full frame
images for both sources in all bands are available in the electronic
version of this paper (Plates 1 to 8).

\section{Results}\label{sec-result}

Figure~\ref{fig-subarray} and \ref{fig-psfsub} show that we do not
detect any point source at the location of Fomalhaut~b indicated by
\citet{Kalas08}. We can however use our PSF roll-subtracted subarray
and full frame images to set limits on the brightness of the planet in
the IRAC bands, and of point sources and extended emission around
Fomalhaut and \epseri.

\subsection{Sensitivity}\label{ssec-sensitivity}

\begin{figure}[t]
\begin{center}
\includegraphics[width=0.7\textwidth,angle=-90]{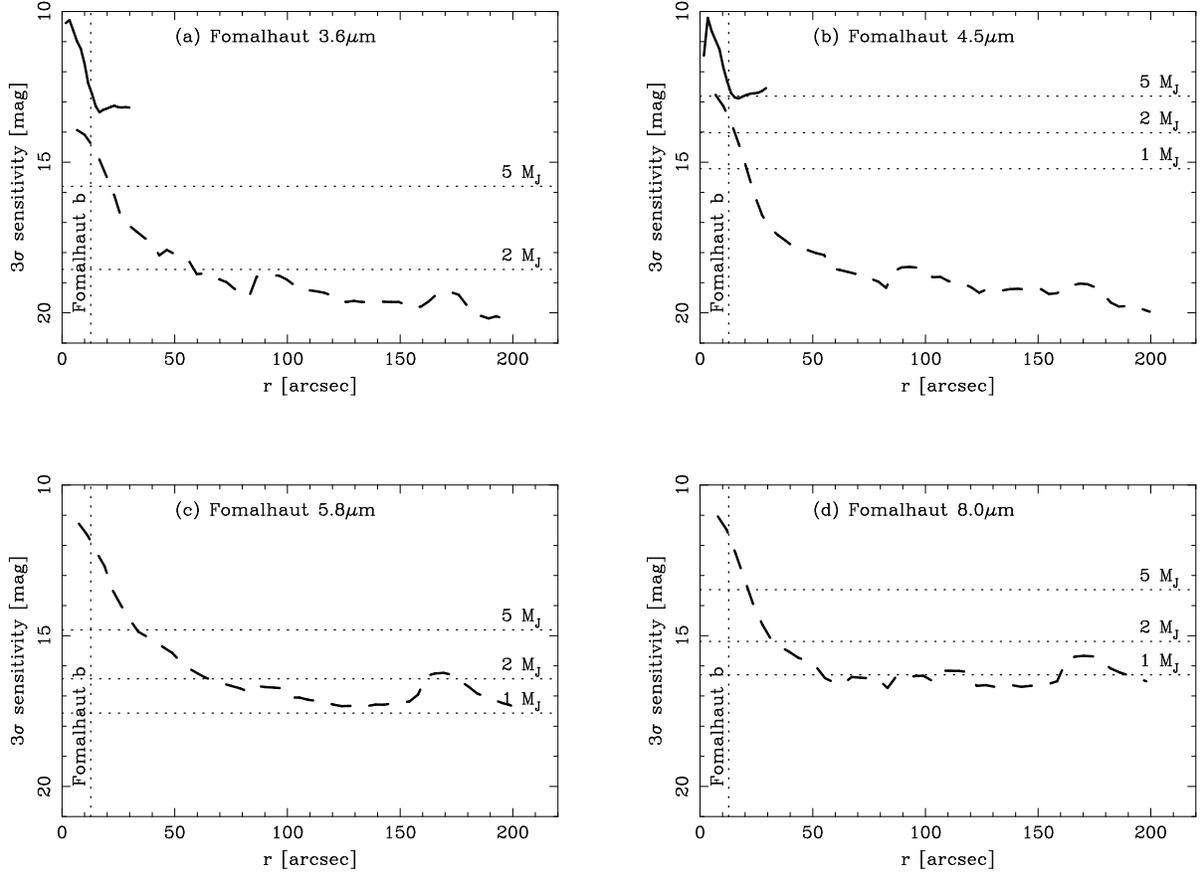}
\end{center}
\caption{ $3 \sigma$ sensitivity of full frame (dashed lines) and
  subarray (solid lines) Fomalhaut images: (a) 3.6~\micron, (b)
  4.5~\micron, (c) 5.8~\micron, and (d) 8.0~\micron. The vertical
  dotted line indicates the projected separation of Fomalhaut b. The
  horizontal lines indicate the magnitudes of 200~Myr planets
  interpolated from 100 and 300~Myr models by
  \citet{Burrows03}. \citet{Fortney08} models predict a similar
  3~M$_J$ limit at 4.5~\micron{}.}\label{fig-fomalhaut-sens}
\end{figure}

\begin{figure}[t]
\begin{center}
\includegraphics[width=0.7\textwidth,angle=-90]{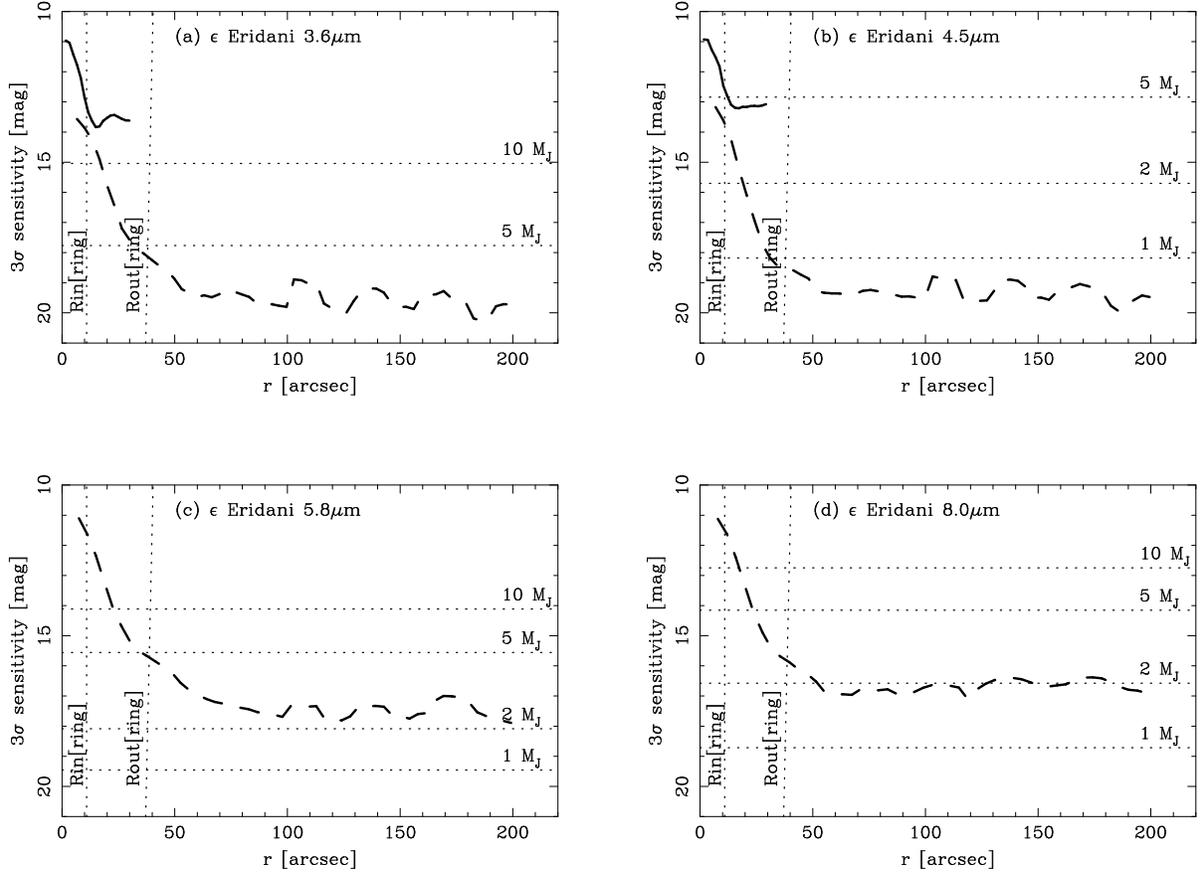}
\end{center}
\caption{
$3 \sigma$ sensitivity of full frame (dashed lines) and
subarray (solid lines) \epseri{} images: (a) 3.6~\micron, (b)
4.5~\micron, (c) 5.8~\micron, and (d) 8.0~\micron. The vertical
dotted lines indicate the inner and outer radii of the submillimeter
ring, and the horizontal lines the magnitudes of 1~Gyr planet models
by \citet{Burrows03}.
}\label{fig-epseri-sens}
\end{figure}

Figure~\ref{fig-fomalhaut-sens} and \ref{fig-epseri-sens} show the
3$\sigma$ radial sensitivity for our subarray and full frame images. We
estimated the sensitivity curves by measuring the root mean square
(RMS) noise in circular annuli having a width $dr \simeq 2
\mathrm{FWHM}(\lambda)$, where $\textrm{FWHM}(\lambda)$ is the Full
Width Half Maximum radius of the IRAC PSF in each band. From the RMS
noise (which is in surface brightness units of MJy/sr) we have then
derived the limiting magnitude in each band, and for each annulus, as
follows: 

\begin{equation}
m_{\textrm{lim}} = -2.5 \log \left[ \frac{3 \pi \left( \textrm{FWHM}/2
   \right)^2 \cdot \textrm{RMS}} {F_0 (r=\textrm{FWHM}/2)} \right] 
\end{equation}

\noindent
where $F_0$ is the flux of the IRAC PSF, normalized as Vega, within a
circular aperture with the diameter of the PSF FWHM, and the factor 3
has been introduced to obtain $3\sigma$ sensitivities. For a Gaussian
noise pattern, these $3\sigma$ sensitivities would guarantee a 99.7\%
detection probability. Given that the PSF-subtraction residual noise
is not Gaussian, however, we have tested these sensitivity curves by
``planting'' point sources of different brightness in the PSF
roll-subtracted images, verifying that the plotted $3\sigma$ curves
indeed guarantee the detection of the sources in all cases.

The two figures also show the expected magnitudes of 1, 2, 5 and
10~M$_J$ planets of 200~Myr and 1~Gyr of age, interpolated from models
by \citet{Burrows03}. Similar estimates can be obtained using the more
recent models by \citet{Fortney08}. The vertical line in
Figure~\ref{fig-fomalhaut-sens} indicates the projected separation of
Fomalhaut~b from \citet{Kalas08}, while the vertical lines in
Figure~\ref{fig-epseri-sens} show the inner and outer radii of the
\epseri{} submillimeter ring (from \citealt{Backman09}).

The plots show that the subarray images are sensitive to point sources
as close as $\sim 3$\asec{} from the central star. The sensitivity at
such small radii is generally poor ($\simeq 10$~mag for Fomalhaut and
$\simeq 11$~mag for \epseri), not sufficient to detect planetary mass
bodies around either stars, but enough to detect T dwarfs. As
explained in Section~\ref{sec-obs} the subarray images sensitivity is
limited by the residual errors from the muxstripe correction, and
``flattens-out'' at radii larger than ~15\asec. The extent of the
muxstripe artifacts in subarray mode was not expected to be so strong
at the time these observations were proposed. For radii larger than
$\simeq 6$\asec, however, our new analysis of the full frame PSF
roll-subtracted images provides a better sensitivity. Between $\sim 6$
to 50\asec{} from the star, our full frame images are limited by the
PSF subtraction noise, while for radii larger than $\sim 50$\asec{}
the sensitivity is limited by the noise in the background.

The $3\sigma$ sensitivity limits at the position of Fomalhaut~b
(Figure~\ref{fig-fomalhaut-sens}) show that the planet has a
brightness lower than 0.5, 0.7, 2.05 and 1.39~mJy at 3.6, 4.5, 5.8 and
8.0~\micron{} respectively (see Table~\ref{tab-fomb}). According to
\citet{Fortney08} this implies a model-dependent mass of $\la 3$~M$_J$
(for a $\sim 200$~Myr age), in agreement with a similar limit inferred
by \citet{Kalas08} from ground-based near-IR data using the same
models. Our images provide the first brightness upper limits for the
planet at 4.5, 5.8, and 8.0~\micron{} (an L' band limit is given in
\citealt{Kalas08}).

From the noise RMS derived above, we can also set limits to the
surface brightness of the debris disks around both stars, listed in
Table~\ref{tab-disk-sens}. Note how our sensitivity at the location of
the Fomalhaut ring NW ansa is as much as 5 times better (1.7
magnitudes) than the sensitivity at the location of the SE ansa. This
is because the SE ansa happens to overlap with high noise residuals
from PSF ``pull-down'' electronic artifacts, locally decreasing our
sensitivity below the circular average level plotted in
Figure~\ref{fig-fomalhaut-sens}. Table~\ref{tab-disk-sens} also shows
our sensitivity limits for the \epseri{} second asteroid belt (at
20~AU radius) and for the inner and outer edge of the sub-millimeter
ring (at 35 and 100~AU respectively) described by \citet{Backman09}.

\citet{Macintosh03} and \citet{Janson08} derive limits for the mass of
planetary mass bodies inside the inner rim of the \epseri{}
sub-millimeter ring of 5 and 3~M$_J$ respectively, compared to our
limit $M \ga 4$~M$_J$ (mass limits estimated for $\sim 1$~Gyr models).
Outside the sub-millimeter ring our limit ($M \ga 1$~M$_J$, 1~Gyr
model age) is superior to any other available observation, including
our previous analysis in \citet{Marengo06}.\footnote{Note that
  Figure~7 in \citet{Marengo06} shows the sensitivity in an area of
  the PSF-subtracted image far from diffraction spikes or other
  artifacts. Figure~\ref{fig-epseri-sens} in the present work shows
  instead a circularly averaged sensitivity, and is superior to the
  equivalent measure performed on our 2006 PSF-subtracted images.}

\subsection{Point Source Photometry}\label{ssec-photometry}

To search for low mass companions around Fomalhaut and \epseri{} we
have measured the photometry of all point sources detected in the PSF
roll-subtracted images of both stars. As shown in
Figure~\ref{fig-fomalhaut-sens} and \ref{fig-epseri-sens} we are
sensitive to substellar objects with mass lower than 1~M$_J$ outside
the debris rings of both stars (from models with age of 200~Myr and
1~Gyr respectively).

To measure the photometry of all the point sources in the field of
view of the full frame images of both stars, we first converted the
data from units of surface brightness flux density to magnitudes, by
dividing by the flux conversion factors found in the image headers of
the BCD, and by multiplying by the effective integration time. The
IRAC absolute calibration is based on observations of standard stars
measured with aperture photometry using a source aperture with a
radius of 10 native IRAC pixels ($\sim 12.2$\asec) in each channel
(this is equivalent to state that the aperture correction of 10 IRAC
pixel radius apertures is exactly 1 according to this convention). The
background was estimated using an annulus centered on the source
position with an inner radius and width of 10 IRAC native pixels
\citep{Reach05}. Because our two targets are located in semi-crowded
fields, we chose to use a smaller source aperture with a radius of
3.6\asec{} in order to avoid contaminating flux from other, nearby
sources. An additional benefit of using a smaller source aperture is
an improved S/N for many of the fainter sources. For the background
estimation, we used the annulus with an inner radius of 4.8\asec{} and
a width of 2.4\asec. Photometry was extracted using the IRAF phot
package.  We determined an aperture correction for these nonstandard
aperture sizes by comparing the photometry from a set of bright stars
in our images using the nominal parameters.  We then determined the
necessary zero points to produce photometry in the standard
calibration defined by \citet{Reach05}. The zero points used are
listed in Table~\ref{tab-cal}.

We have first analyzed the colors of all sources detected in multiple
IRAC bands. Given the lower sensitivity at 5.8 and 8.0~\micron, and
the CH$_4$ absorption expected at 3.6~\micron{} for T-dwarfs and gas
giant planets, this limits our discovery space to objects with a
200~Myr model dependent mass of $\sim 2$~M$_J$ around Fomalhaut, and
$\sim 4$~M$_J$ around \epseri{} (1~Gyr model age). We used the same
$k$-Nearest Neighbor method search technique based on the colors and
absolute magnitudes of brown dwarf and planet models we adopted in
\citet{Marengo06}, described in detail in \citet{Marengo09}. None of
the sources detected in all four bands of full array imaging had the
colors and absolute magnitudes expected for sub-stellar mass objects.

All sources that were detected at 3.6 and 4.5~\micron, but missed at
5.8 and/or 8.0~\micron, have also been rejected. Based on
\citet{Burrows03} and \citet{Fortney08} models none of them possess a
[3.6]$-$[4.5] color red enough to be a planetary mass object for their
measured 4.5~\micron{} magnitude.

We then focused on sources detected at 4.5~\micron{} but not detected
at 3.6~\micron: these sources can be either very red extragalactic
sources \citep{Huang04, Barmby04} or methane dwarfs and planets. Red,
mass losing background giants (such as Asymptotic Giant Branch stars)
have positive [4.5]$-$[8.0] colors \citep{Marengo08}, and are thus
distinguishable from T dwarfs and planets that have instead a
8.0~\micron{} flux equal or lower than the 4.5 micron
flux. Table~\ref{tab-fdrop} and \ref{tab-eedrop} lists all
3.6~\micron{} ``dropout'' sources for Fomalhaut and \epseri{}
respectively. Note how several of them can be discarded because their
flux at 5.8 and 8.0~\micron{} makes them likely to be background mass
losing giants. Comparison with other datasets could enable the
rejection of other dropout sources based on (lack of) common proper
motion with the primary star. To our knowledge, however, none of the
dropout sources in Table~\ref{tab-fdrop} and \ref{tab-eedrop} has been
detected in any other observation at optical or infrared wavelength.

\begin{figure}[t]
\begin{center}
\includegraphics[width=0.6\textwidth,angle=0]{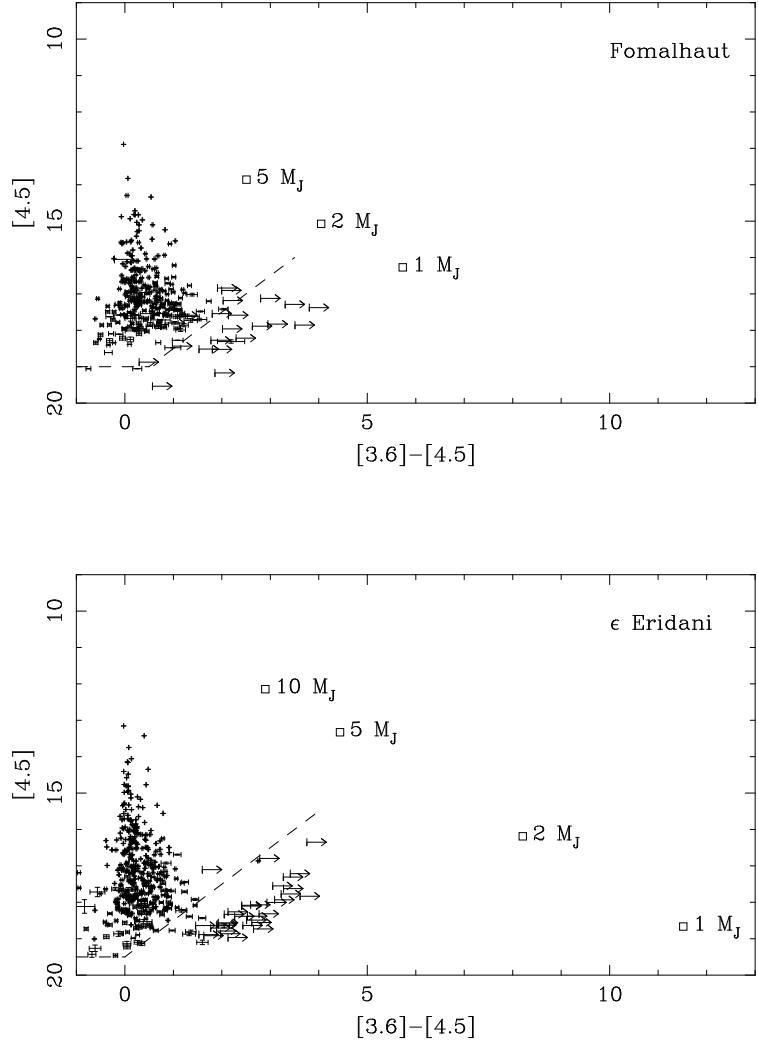}
\end{center}
\caption{ Color-magnitude diagram of all point sources detected in the
  PSF roll-subtracted 4.5~\micron{} image of Fomalhaut (top) and
  \epseri{} (bottom). Points with error bars are sources detected in
  both 3.6 and 4.5~\micron{} IRAC bands, while the arrows indicate
  sources that are undetected at 3.6~\micron{} (for which we used the
  image sensitivity at the source location as 3.6~\micron{}
  limit). The squares indicate the colors and magnitudes of planets
  from \citet{Burrows03} models, interpolated to the age of each
  star. The dashed line indicates the average $3 \sigma$ sensitivity
  limit of our 3.6 and 4.5~\micron{} images.  }
\label{fig-phot}
\end{figure}

Figure~\ref{fig-phot} shows the [3.6]$-$[4.5] vs. [4.5]
color-magnitude diagram of all sources detected in both IRAC bands 1
and 2 (data points with error-bars) and of all 3.6~\micron{} dropout
sources (arrows). The [3.6]$-$[4.5] color of the dropout sources is a
lower limit, estimated using the local 3.6~\micron{} $3\sigma$
sensitivity derived in Section~\ref{ssec-sensitivity}. The dropout
sources within the Fomalhaut field of view have 4.5~\micron{}
magnitudes expected for $\la 1$~M$_J$ 200~Gyr planets, according to
\citet{Burrows03} models. However, their [3.6]$-$[4.5] colors are
still compatible with the colors of red extragalactic sources
(e.g. see \citealt{Stern07}). Similarly, the dropout sources in the
\epseri{} field (that, if they were $\sim 1$~Gyr planets, would have a
mass $\la 2$~M$_J$) also have color limits compatible with being
background galaxies.

\begin{figure}[t]
\begin{center}
\includegraphics[width=0.7\textwidth]{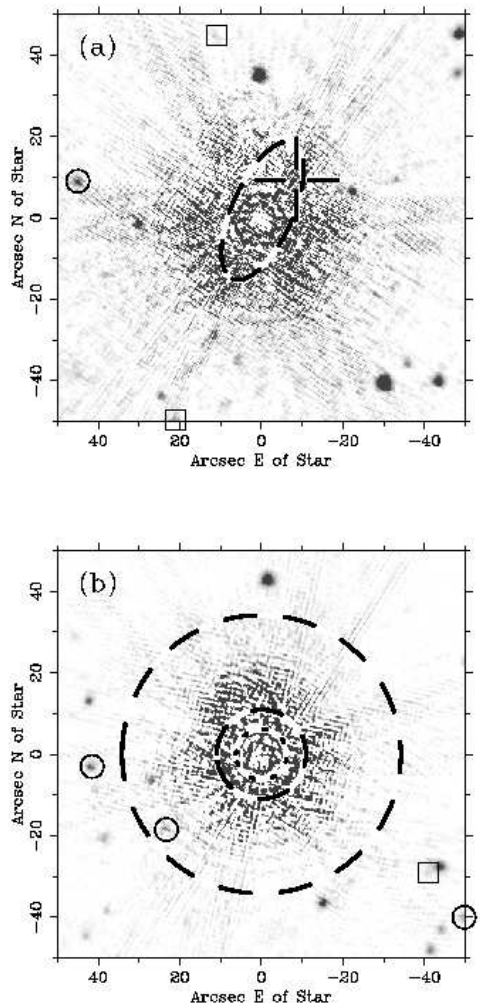}
\end{center}
\caption{Final 4.5~\micron{} PSF roll-subtracted full frame
  images. Top panel shows the inner 50$\times$50\asec{} ($\sim
  385$$\times$385~AU) of Fomalhaut. The ellipse indicates the location
  of the debris ring, and the cross the position of Fomalhaut b. The
  bottom panel shows the same area (in arcsec, corresponding to $\sim
  160$$\times$160~AU) around \epseri. The dotted circle indicates the
  position of the 20~AU asteroid belt, while the two dashed circles
  indicate the inner and outer boundaries of the sub-millimeter ring
  as derived by \citet{Backman09}. The images are scaled linearly from
  0 to 0.3~MJy/sr in the surface brightness color scale. Circle points
  mark the position of 3.6~\micron{} ``dropout'' sources that are
  likely background mass-losing giants while square points indicate
  sources detected only at 4.5~\micron.}\label{fig-zoom}
\end{figure}

Figure~\ref{fig-zoom} shows the position of the 3.6~\micron{} dropout
sources within 50$\times$50\asec{} from Fomalhaut and \epseri, and
their relative position with respect to the debris disks. This area
corresponds to $\sim 385$$\times$385~AU at the distance of Fomalhaut,
and $\sim 160$$\times$160~AU at the distance of \epseri. Three dropout
sources are detected within 300--400~AU from the Fomalhaut ring, but
none of them is in close proximity, or inside, the debris ring. Of
these three sources, one has [4.5]$-$[5.8] and [4.5]$-$[8.0] color
larger than $\sim 1.5$, which is characteristic of background mass
losing giants. The other two sources are companion candidates with
mass lower than 1~M$_J$ (according to 200~Myr \citealt{Burrows03}
models). Of the four dropout sources detected near \epseri, one is
inside the sub-millimeter ring. This source, as well as two of the
sources detected within $\sim 100$~AU from the outer radius of the
\epseri{} sub-millimeter ring, is likely a mass losing giant in the
background. The remaining source, $\sim 90~AU$ from the outer rim of
the debris disk, is a planet candidate with a mass as low as 2~M$_J$
(based on 1~Gyr \citealt{Burrows03} models).

Of all the dropout sources listed in Table~\ref{tab-eedrop}, only 5
were detected at 4.5~\micron{} in \citet{Marengo06}, due to the lower
quality of the PSF subtraction in that work. All the other
3.6~\micron{} dropout sources found in \citet{Marengo06}, Table~3,
have been detected above $3\sigma$ in our current PSF roll-subtracted
images, and have been excluded as planet candidates. Thanks to our
better sensitivity at 4.5~\micron, we detect 3 sources that were
missed by \citet{Macintosh03} in their field of view. Of these
sources, two are 3.6~\micron{} dropouts (shown in
Figure~\ref{fig-zoom}, E of \epseri). The colors of these sources
suggest that they are background stars or galaxies.

\section{Discussion and Conclusions}\label{sec-disc}

\begin{figure}[t]
\begin{center}
\includegraphics[width=0.7\textwidth,angle=-90]{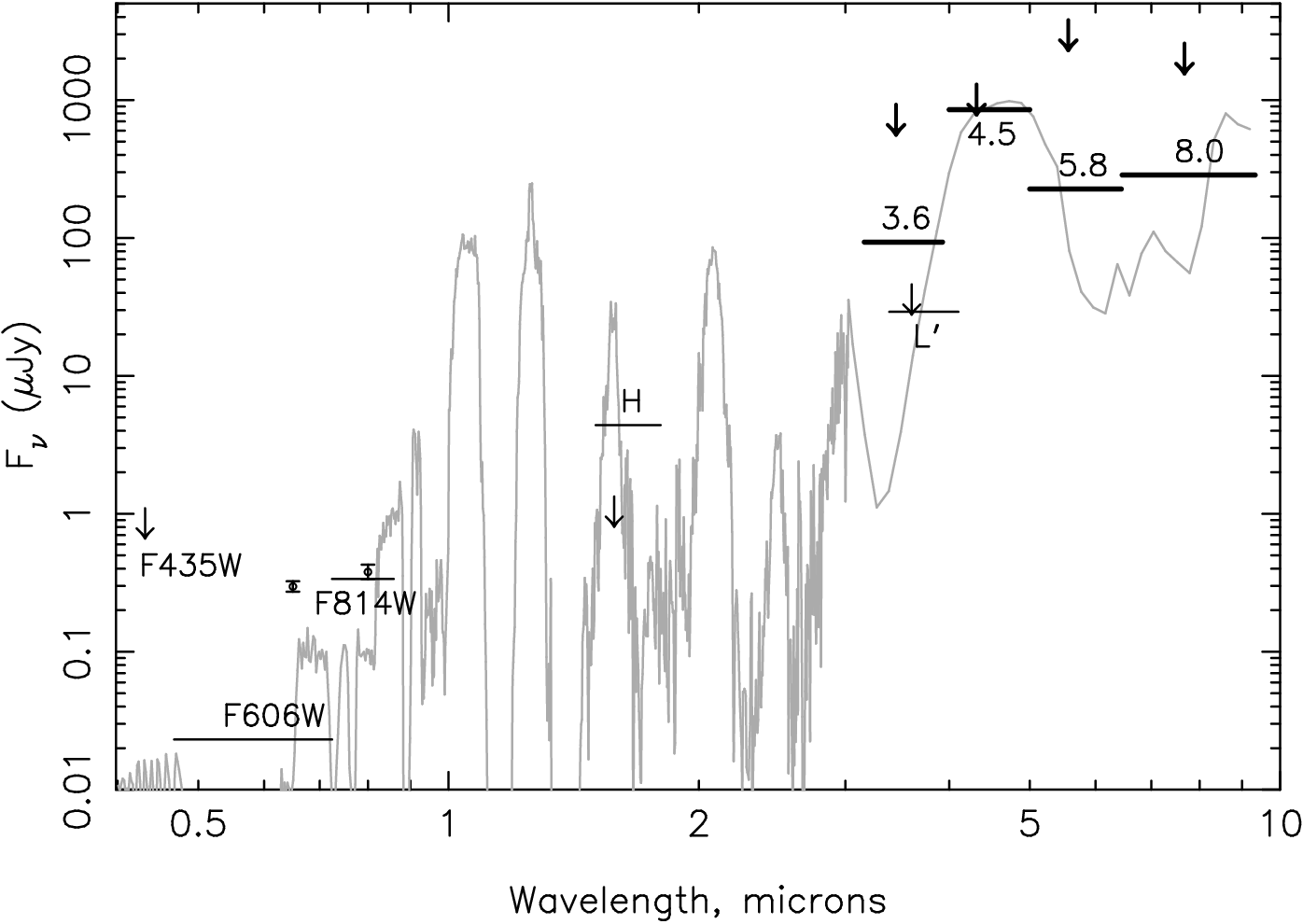}
\end{center}
\caption{Fomalhaut b spectral energy distribution, compared with the
  \citet{Fortney08} model used by \citet{Kalas08} to fit their
  data. Optical photometry and 3$\sigma$ limits are from
  \citet{Kalas08}. Limits are indicated by the tip of the arrows. IRAC
  3$\sigma$ limits (thick arrows) are from Table~\ref{tab-fomb} in
  this work. The horizontal bars mark the equivalent broad-band flux
  found by integrating the model spectrum over the instrumental
  pass-bands. Our 4.5~\micron{} limit, together with the
  \citet{Kalas08} near-IR limits, suggest a Fomalhaut~b mass lower
  than 3~M$_J$ (for a 200~Myr age model), and is inconsistent with the
  measured optical flux of the planet.}\label{fig-fomalhaut-sed}
\end{figure}

The physical origin of the optical light detected by \citet{Kalas08}
from Fomalhaut b is unclear.  \citet{Fortney08} model of the thermal
emission from a 400 K object of $\sim$ 1.2 Jupiter radius (corresponding
to a 3 M${J}$ object at age 200 Myr) can reproduce the 0.8~\micron{}
flux density observed by HST.  This model is shown in
Figure~\ref{fig-fomalhaut-sed}, along with the \spitzer{} mid-infrared
upper limits from this study and the photometry from \citet{Kalas08}.
As already noted by the latter authors, this model predicts a
1.6~\micron{} brightness $\sim 5$$\times$ larger than the observed
upper limit, and a 0.6~\micron{} flux density orders of magnitude
fainter than the observed HST detection (even taking into account the
observed 0.6~\micron{} variability, possibly related to variable
H$\alpha$ emission from a hot planetary chromosphere).  Reconciling
the thermal emission model to the observations would require an
additional physical mechanism to produce the 0.6~\micron{} emission,
and that atmosphere models be revised to suppress the expected
1.6~\micron{} emission.  The \spitzer{} 4.5~\micron{} upper limit lies
on top of this thermal emission model, and thus strengthens these
conclusions.  Any revision of the thermal emission model to account
for the 1.6~\micron{} non-detection \emph{is constrained} by our
results.  In particular, our 4.5~\micron{} upper limit does not allow
a lot of room for the suppressed 1.6~\micron{} luminosity to emerge
instead through the largest low-opacity spectral window in
methane-dominated atmospheres.  The strength of this constraint can
only be evaluated through new model atmosphere work directed toward
finding a thermal emission solution to the properties of Fomalhaut b.
In the interim, reflection from a circumplanetary dust disk remains
the simplest model to explain the observed fluxes from the object.

Our surface brightness upper limits (Table~\ref{tab-disk-sens}) for
the Fomalhaut ring can be compared to the contrast seen in optical
scattered light by \citet{Kalas08}.  On the NW ansa, our 3.6~\micron{}
limiting surface brightness (in mag/arcsec$^2$) is 17.5 mag fainter
than the star itself.  At 0.6~\micron{}, the HST detected surface
brightness of 21.0 mag/arcsec$^2$ is 19.9 mag fainter than the star.
To escape detection, the V-L' color of the ring must therefore be less
than 2.4 mag. V-L' colors of debris disks have however still not been
measured, so the strength of our color constraint is unclear.
However, the reddest class of Kuiper Belt objects in our solar system
has V-K colors less than 1.5 mag \citep{Cruikshank07}.  If the
Fomalhaut dust has similar properties, it thus should not have been
detected in our IRAC images.

Similarly, the expected surface brightness of the \epseri{}
sub-millimeter ring, as derived by \citet{Backman09}, is at least 1 or
2 orders of magnitude below our sensitivity, in agreement with the
0.011~MJy/sr \citet{Proffitt04} limit set by STIS camera observations
in scattered light. According to the \citet{Backman09} model, the
\epseri{} 20~AU asteroid belt is expected to have a surface brightness
of $\sim 21.7$ mag/arcsec$^2$ ($\sim 0.3$~MJy/sr) in the V band,
corresponding to $\sim 0.02$~MJy/sr at 3.6~\micron. This is also well
below the sensitivity limits listed in Table~\ref{tab-disk-sens} and
cannot be detected in our IRAC images.

None of the 3.6~\micron{} ``dropout'' sources listed in
Table~\ref{tab-fdrop} and \ref{tab-eedrop} have colors and magnitudes
strongly suggesting that they are planetary mass companions of
Fomalhaut or \epseri, rather than red background objects (even though
repeated observations aimed to measure common proper motion are
required to clarify the issue). Our detection limits for point sources
outside the debris disks of Fomalhaut and \epseri{} thus imply the
likely absence of any widely separated companions of the two stars
with mass larger than 1~M$_J$ (according to \citealt{Burrows03} models
of 200~Myr and 1~Gyr age, respectively). Inside the rings (and down to
the 3--6\asec{} saturation radius of our images) our limits obtained
with the \spitzer{} 85~cm aperture telescope are comparable, or
superior, to model-dependent mass detection limits from infrared
observations obtained at 8-10~m class ground-based telescopes.

\acknowledgements

We thank John Krist and James Graham for their advice on PSF subtractions
and the application of the Fortney atmosphere models to our dataset.
This work is based on observations made with the \spitzer{} Space Telescope, 
which is operated by the Jet Propulsion Laboratory, California Institute 
of Technology under a contract with NASA.  Support for this work was provided 
by NASA under \spitzer{} General Observer grants 30754 to the 
Harvard-Smithsonian Center for Astrophysics and to JPL.



\clearpage
\newpage
\begin{deluxetable}{llrrrl}
\tablecaption{Observing Log\label{tab-obs}}
\tablewidth{\textwidth}
\tabletypesize{\footnotesize}
\tablehead{
\colhead{Source} & \colhead{Mode} & \colhead{AOR KEY} &
\colhead{Obs. Date} & \colhead{Tot. Exp.} & \colhead{Wavelengths}\\
& & & \colhead{[JD-2400000.5]} & \colhead{[sec]} &
\colhead{[\micron]} }
\startdata
Fomalhaut & Full     &  4875776 & 53337.306 &  499.2 & 3.6, 4.5, 5.8, 8.0 \\
Fomalhaut & Full     &  9015040 & 53309.253 &  499.2 & 3.6, 4.5, 5.8, 8.0 \\
Fomalhaut & Subarray & 18951936 & 54063.253 & 1468.8 & 3.6, 4.5 \tablenotemark{(a)} \\
Fomalhaut & Subarray & 18952192 & 54095.207 & 1468.8 & 3.6, 4.5 \\
\epseri   & Full     &  4876032 & 53013.836 & 3369.6 & 3.6, 4.5, 5.8, 8.0 \\
\epseri   & Full     &  4876288 & 53052.781 & 3369.6 & 3.6, 4.5, 5.8, 8.0 \\
\epseri   & Subarray & 18951424 & 54147.336 & 1468.8 & 3.6, 4.5 \\
\epseri   & Subarray & 18951680 & 54356.199 & 1468.8 & 3.6, 4.5 \\
\enddata
\tablenotetext{a}{Total Exposure time for the 4.5~\micron{} image is
  1419.5~s} 
\end{deluxetable}

\clearpage
\newpage
\begin{deluxetable}{lrrrr}
\tablecaption{Source PSF-fitting Photometry\label{tab-psf-phot}}
\tablehead{
\colhead{Source} & \colhead{[3.6]} & \colhead{[4.5]} 
                & \colhead{[5.8]} & \colhead{[8.0]}
}
\startdata
Vega\tablenotemark{a} &    0.00 &    0.00 &    0.00 &    0.00 \\
Fomalhaut             &    0.98 &    0.98 &    0.98 &    0.98 \\
\epseri               &    1.60 &    1.63 &    1.61 &    1.59 \\
$\epsilon$~Indi       &    2.10 &    2.15 &    2.12 &    2.07 \\
Sirius                & $-$1.38 & $-$1.38 & $-$1.38 & $-$1.35 \\
\enddata
\tablenotetext{a}{Used as reference, it has 0 Vega-magnitude by
  definition} 
\end{deluxetable}

\clearpage
\newpage
\begin{deluxetable}{lcccc}
\tablecaption{IRAC Photometric Calibration\tablenotemark{1}\label{tab-cal}}
\tablewidth{1\textwidth}
\tablehead{ & \multicolumn{4}{c}{IRAC band}\\
\colhead{Item} &
\colhead{3.6~\micron} &
\colhead{4.5~\micron} &
\colhead{5.8~\micron} &
\colhead{8.0~\micron} }
\startdata
Isophotal $\lambda$ [\micron]  & 3.550  & 4.493  & 5.731  & 7.872  \\
FLUXCONV [(MJy/sr)/(DN/s)]     & 0.1088 & 0.1388 & 0.5952 & 0.2021 \\
GAIN [e/DN]                    & 3.3    & 3.71   & 3.8    & 3.8    \\
Zero point magnitudes\tablenotemark{2}
                              & 16.981 & 16.512 & 16.013 & 15.439 \\
$F_\nu$(Vega) [Jy]             & 280.9  & 179.7  & 115.0  & 64.1   \\
\enddata
\tablenotetext{1}{Based on the IRAC Data Handbook ver. 3.0 (2006)}
\tablenotetext{2}{For pixel size 0.24\asec/pix, include aperture
 correction for 3.6\asec{}  aperture and sky annulus with 4.8\asec{}
 and 7.2\asec{}  inner and outer radii}
\end{deluxetable}

\clearpage
\newpage
\begin{deluxetable}{lcccc}
\tablecaption{3$\sigma$ limiting Sensitivity at the Fomalhaut b 
Radius\label{tab-fomb}}
\tablewidth{1\textwidth}
\tablewidth{1\textwidth}
\tablehead{
\colhead{Mode} & 
\colhead{$F_{lim}$(3.6\micron)} &
\colhead{$F_{lim}$(4.5\micron)} &
\colhead{$F_{lim}$(5.8\micron)} &
\colhead{$F_{lim}$(8.0\micron)} \\
& \colhead{[mJy]} & \colhead{[mJy]} &
  \colhead{[mJy]} & \colhead{[mJy]}
}
\startdata
Full frame & 0.50 & 0.70 & 2.05 & 1.39 \\
Subarray   & 2.54 & 1.83 & \nodata & \nodata \\
\enddata
\end{deluxetable}

\clearpage
\newpage
\begin{deluxetable}{lrrrr}
\tablecaption{Fomalhaut and \epseri{} Disks brightness
 limits\label{tab-disk-sens}} 
\tablewidth{1\textwidth}
\tablewidth{1\textwidth}
\tabletypesize{\footnotesize}
\tablehead{
\colhead{Mode} & 
\colhead{$S_{lim}$(3.6\micron)} &
\colhead{$S_{lim}$(4.5\micron)} &
\colhead{$S_{lim}$(5.8\micron)} &
\colhead{$S_{lim}$(8.0\micron)} \\
& \colhead{[MJy/sr]} & \colhead{[MJy/sr]} &
  \colhead{[MJy/sr]} & \colhead{[MJy/sr]}
}
\startdata
Fomalhaut ring (NW ansa)         & 1.19 & 1.43 &  3.84 & 2.10 \\
Fomalhaut ring (SE ansa)         & 5.18 & 5.94 & 20.26 & 3.39 \\
\epseri{} asteroid belt 2        & 2.25 & 2.07 &  6.98 & 2.72 \\
\epseri{} sub-mm ring inner edge & 1.52 & 1.20 &  3.96 & 1.69 \\
\epseri{} sub-mm ring outer edge & 0.04 & 0.02 &  0.11 & 0.04 \\
\enddata
\end{deluxetable}

\clearpage
\newpage
\begin{deluxetable}{rrrcccl}
\tablecaption{Fomalhaut 3.6~\micron{} Dropout Sources\label{tab-fdrop}}
\tablewidth{1\textwidth}
\tablewidth{1\textwidth}
\tablehead{
 \colhead{RA[2000]} & \colhead{Dec[2000]} & \colhead{D [AU]} &
 \colhead{[4.5]} & \colhead{[5.8]} & \colhead{[8.0]} & \colhead{Notes}
}
\startdata
344.391700 & -29.672331 & 1391 & 19.17$\pm$ 0.17 &  \nodata  &  \nodata  & \\
344.441542 & -29.660608 & 1074 & 17.83$\pm$ 0.04 &  \nodata  &  \nodata  & \\
344.441542 & -29.660608 & 1074 & 17.83$\pm$ 0.04 &  \nodata  &  \nodata  & \\
344.372933 & -29.645378 &  683 & 17.96$\pm$ 0.04 &  \nodata  &  \nodata  & \\
344.413842 & -29.646267 &  662 & 18.27$\pm$ 0.06 &  \nodata  &  \nodata  & \\
344.399342 & -29.641753 &  543 & 19.54$\pm$ 0.00 &  \nodata  &  \nodata  & \\
344.424521 & -29.646136 &  662 & 18.52$\pm$ 0.15 &  \nodata  &  \nodata  & \\
344.382887 & -29.636044 &  421 & 18.43$\pm$ 0.00 &  \nodata  &  \nodata  & \\
344.424175 & -29.644953 &  629 & 17.29$\pm$ 0.05 &  \nodata  &  \nodata  & \\
344.383133 & -29.629925 &  278 & 17.59$\pm$ 0.00 &  \nodata  &  \nodata  & \\
344.419842 & -29.635978 &  379 & 17.18$\pm$ 0.07 &  \nodata  & 17.93$\pm$ 0.45 & \\
344.427429 & -29.619778 &  113 & 16.05$\pm$ 0.00 & 14.70$\pm$ 0.04 &
14.81$\pm$ 0.02 & (1) \\
344.479287 & -29.629350 &  449 & 17.59$\pm$ 0.02 &  \nodata  &  \nodata  & \\
344.433550 & -29.617836 &  177 & 18.98$\pm$ 0.00 &  \nodata  &  \nodata  & \\
344.400317 & -29.607925 &  407 & 17.54$\pm$ 0.07 &  \nodata  &  \nodata  & \\
344.435804 & -29.613942 &  272 & 17.80$\pm$ 0.00 &  \nodata  &  \nodata  & \\
344.416708 & -29.609828 &  348 & 18.87$\pm$ 0.00 &  \nodata  &  \nodata  & \\
344.421133 & -29.604700 &  492 & 17.12$\pm$ 0.04 &  \nodata  &
15.69$\pm$ 0.04 & (1) \\
344.419792 & -29.600181 &  616 & 17.89$\pm$ 0.05 &  \nodata  &  \nodata  & \\
344.418371 & -29.601414 &  581 & 18.52$\pm$ 0.08 &  \nodata  &  \nodata  & \\
344.413029 & -29.574408 & 1329 & 18.22$\pm$ 0.05 &  \nodata  &
17.25$\pm$ 0.21 & (1) \\
344.423208 & -29.568053 & 1506 & 17.86$\pm$ 0.04 &  \nodata  &
16.22$\pm$ 0.06 & (1) \\
\enddata
\tablenotetext{(1) }{[4.5]$-$[5.8] or [4.5]$-$[8.0] color suggest source
being a background giant}
\end{deluxetable}

\clearpage
\newpage
\begin{deluxetable}{rrrcccl}
\tablecaption{\epseri{} 3.6~\micron{} Dropout Sources\label{tab-eedrop}}
\tablewidth{1\textwidth}
\tablewidth{1\textwidth}
\tablehead{
 \colhead{RA[2000]} & \colhead{Dec[2000]} & \colhead{D [AU]} &
 \colhead{[4.5]} & \colhead{[5.8]} & \colhead{[8.0]} & \colhead{Notes}
}
\startdata
53.200488 &  -9.512958 &  724 & 18.32$\pm$ 0.02 & 16.13$\pm$ 0.01 &
18.20$\pm$ 0.13 & (1) \\
53.194846 &  -9.487653 &  541 & 18.71$\pm$ 0.04 &  \nodata  &
16.45$\pm$ 0.06 & (1) \\
53.210854 &  -9.514525 &  690 & 17.93$\pm$ 0.01 & 16.79$\pm$ 0.01 &
16.05$\pm$ 0.00 & (1) \\
53.206725 &  -9.482892 &  402 & 18.55$\pm$ 0.03 &  \nodata  &  \nodata  & \\
53.213421 &  -9.489692 &  417 & 18.79$\pm$ 0.03 &  \nodata  &  \nodata  & \\
53.218321 &  -9.494831 &  448 & 18.89$\pm$ 0.04 &  \nodata  &  \nodata  & \\
53.219821 &  -9.506739 &  574 & 18.00$\pm$ 0.02 &  \nodata  &  \nodata  & \\
53.219896 &  -9.509417 &  604 & 18.73$\pm$ 0.04 & 15.57$\pm$ 0.01 &
16.80$\pm$ 0.04 & (1) \\
53.193787 &  -9.440481 &  479 & 18.49$\pm$ 0.03 &  \nodata  &  \nodata  & \\
53.202425 &  -9.455889 &  335 & 18.64$\pm$ 0.03 &  \nodata  &  \nodata  & \\
53.219771 &  -9.481456 &  299 & 18.38$\pm$ 0.04 &  \nodata  &
\nodata  & (2) \\
53.223296 &  -9.488964 &  366 & 18.92$\pm$ 0.05 & 17.48$\pm$ 0.06 &
\nodata  & (1,2) \\
53.213246 &  -9.447133 &  245 & 18.56$\pm$ 0.04 &  \nodata  &
\nodata  & (2) \\ 
53.219871 &  -9.466025 &  160 & 16.35$\pm$ 0.01 &  \nodata  &  \nodata  & \\
53.217404 &  -9.469114 &  204 & 17.21$\pm$ 0.02 &  \nodata  &
15.67$\pm$ 0.03 & (1) \\
53.233725 &  -9.477178 &  220 & 17.55$\pm$ 0.02 &  \nodata  &  \nodata  & \\
53.202654 &  -9.420789 &  543 & 18.10$\pm$ 0.02 &  \nodata  &  \nodata  & \\
53.209637 &  -9.429756 &  413 & 18.86$\pm$ 0.04 &  \nodata  &  \nodata  & \\
53.241296 &  -9.471792 &  193 & 18.39$\pm$ 0.08 &  \nodata  &  \nodata  & \\
53.239879 &  -9.477864 &  246 & 18.59$\pm$ 0.06 &  \nodata  &  \nodata  & \\
53.209096 &  -9.416711 &  543 & 18.08$\pm$ 0.02 &  \nodata  &  \nodata  & \\
53.225883 &  -9.443644 &  179 & 18.63$\pm$ 0.08 &  \nodata  &  \nodata  & \\
53.243187 &  -9.458864 &  135 & 16.79$\pm$ 0.03 & 16.57$\pm$ 0.14 &
15.28$\pm$ 0.02 & (1,3,4) \\
53.238017 &  -9.463064 &   93 & 17.11$\pm$ 0.10 & 16.25$\pm$ 0.14 &
16.42$\pm$ 0.12 & (1,4) \\
53.247821 &  -9.466464 &  211 & 18.07$\pm$ 0.04 & 17.99$\pm$ 0.12 &
17.09$\pm$ 0.07 & (1) \\
53.258658 &  -9.484592 &  436 & 18.33$\pm$ 0.02 &  \nodata  &
\nodata  & (2) \\
53.250783 &  -9.458222 &  222 & 17.31$\pm$ 0.02 & 16.73$\pm$ 0.04 &
15.69$\pm$ 0.02 & (1) \\
53.229712 &  -9.404978 &  613 & 18.64$\pm$ 0.01 &  \nodata  &  \nodata  & \\
53.245933 &  -9.422414 &  444 & 18.26$\pm$ 0.02 &  \nodata  &  \nodata  & \\
53.256567 &  -9.436811 &  379 & 17.77$\pm$ 0.02 &  \nodata  &  \nodata  & \\
53.255775 &  -9.436803 &  373 & 17.62$\pm$ 0.02 &  \nodata  &  \nodata  & \\
53.283862 &  -9.494122 &  731 & 18.97$\pm$ 0.02 & 15.20$\pm$ 0.00 &
\nodata  & (1) \\
53.273888 &  -9.428683 &  595 & 17.83$\pm$ 0.01 & 16.35$\pm$ 0.01 &
15.87$\pm$ 0.00 & (1) \\
\enddata
\tablenotetext{(1) }{[4.5]$-$[5.8] or [4.5]$-$[8.0] color suggest source
being a background giant}
\tablenotetext{(2) }{Source listed in Table~3, \citet{Marengo06}}
\tablenotetext{(3) }{Source 11 in Table~4, \citet{Marengo06}}
\tablenotetext{(4) }{Within field of view, but not detected by
 \citet{Macintosh03}}
\end{deluxetable}


\clearpage
\newpage
\onecolumn
%

\clearpage
\newpage
\onecolumn
%

\clearpage
\newpage
\onecolumn
%

\clearpage
\newpage
\onecolumn
%

\clearpage
\newpage
\onecolumn
%

\clearpage
\newpage
\onecolumn
%

\clearpage
\newpage
\onecolumn
%

\clearpage
\newpage
\onecolumn
%


\end{document}